\documentclass{article}
\usepackage[english]{babel}
\usepackage{graphicx} 
\usepackage[style=ieee, sorting=none]{biblatex}
\usepackage{csquotes}
\usepackage{authblk}

\title{Visual acuity and contrast sensitivity under monochromatic yellow light}
\usepackage{authblk}
\author[1]{Elisabetta Baldanzi}
\author[2]{Paolo Antonino Grasso}
\author[3]{Clara Gori}
\author[2]{Massimo Gurioli}
\author[2]{Federco Tommasi}
\author[1]{Alessandro Farini}
\affil[1]{CNR-Istituto nazionale di Ottica, Firenze, Italy}
  \affil[2]{Dipartimento di Fisica e Astronomia, Università di Firenze, via Sansone 1, 50019 Sesto Fiorentino, Italy}
  \affil[3]{Corso di Studi in Ottica e Optometria, Università degli Studi di Firenze, Firenze, Italy}

\date{January 22, 2024}

\begin{document}

\maketitle

\begin{abstract}
This study investigates the impact of monochromatic lighting on visual acuity (VA) and contrast sensitivity (CS). 
Traditional assessments of VA and CS are typically conducted under the illumination of ``white'' light, but variations in color temperature can influence outcomes. 
Utilizing data from an exhibition by Olafur Eliasson, where a room was illuminated with low-pressure sodium lamps, creating an almost monochromatic yellow light, we compared visual assessments in the yellow room with conventional lighting in a white room.

For VA, the results show no significant differences between the two lighting conditions, while for CS, a more nuanced situation is observed. 
The bias in CS measurements is clinically relevant, and the p-value suggests that further investigation with a larger, more diverse sample may be worthwhile.

Despite limitations, such as higher illumination conditions than standard protocols, the unique ``laboratory'' offered by the exhibition facilitated measurements not easily achievable in a traditional setting.
\end{abstract}

\section{Introduction}
Assessing visual acuity (VA) and contrast sensitivity (CS) is a routine practice conducted in standard lighting conditions. Specifically, both VA and CS are typically examined under the illumination of ``white'' light. However, it is important to note that the term ``white'' light might be overly generic, since variations in color temperature can influence the outcomes. 
For instance, diverse color temperatures of light may result in distinct VA values, highlighting the need for a nuanced consideration of lighting conditions in visual assessments\cite{amouzadeh2023impact}. 

If even a slight variation in color temperature has the potential to influence visual acuity and contrast sensitivity  outcomes, the degree of variability in results may be heightened in situations where lighting significantly deviates from white light. 

Probably, the earliest experiment on visual acuity with monochromatic light was conducted by Luckiesh in 1911 \cite{luckiesh_monochromatic_1911}. 
Luckiesh utilized a green monochromatic light emitted by a mercury vapor lamp and a tungsten lamp equipped with a green bandpass filter. 
The bandwidth of the filter significantly exceeded the width of the mercury emission spectrum. 
Luckiesh observed that, to achieve the same visual acuity, the filtered tungsten 

Subsequently, with the availability of sodium-vapor lamps, the experiment was revisited, this time utilizing the yellow line of sodium \cite{luckiesh_visual_1933}. 
It was noted that monochromatic light indeed resulted in enhanced visual acuity, although the degree of improvement diminished as brightness increased.

Recently, experiments were conducted utilizing LED lighting\cite{ramamurthy2004determining}; however, the statistical significance of the results remains unclear.

\begin{figure}[htbp]
    \centering
    \includegraphics[width=\linewidth]{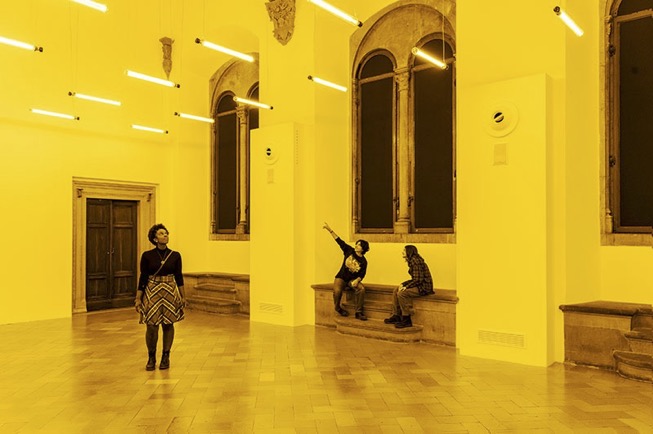}
    \caption{``Room for one color'' . Photo Ela Bialkowska, OKNO Studio. Courtesy Fondazione Palazzo Strozzi, Florence ©2022 Olafur Eliasson.}
    \label{fig:salagialla}
\end{figure}

This work stems from a fortunate opportunity. 
In Florence, from September 22, 2022, to January 22, 2023, the exhibition ``Olafur Eliasson: nel tuo tempo'' took place (the title was not translated into English by the artist's choice), dedicated to the renowned contemporary artist. One of the artworks in this exhibition was ``room for one color'', where the artist illuminated an entire room with low-pressure sodium lamps, creating an almost monochromatic yellow illumination (Fig.\ref{fig:salagialla}).
Since the Fondazione Strozzi, the organizer of the exhibition, allowed us to conduct guided tours for professors and students of the Optics and Optometry degree program, we noticed that many people claimed to ``see better'' inside that room. 
This sparked our desire to investigate whether this qualitative observation could be measured, and, thanks to the Foundation, we were able to conduct measurements within the exhibition hall and, for comparison, in a room with conventional lighting.

The improved vision within the yellow room could be explained in terms of two different phenomena: the reduction of chromatic aberration of the eye and the filtering effect of blue light on the achromatic and chromatic channels of visual perception.

Dispersion of the refracting medium in human eyes results in longitudinal chromatic aberration (LCA). 
Given that  LCA induces significant defocus, correcting this aberration additionally should enhance visual quality. 
Unlike monochromatic aberrations, the LCA of the eye is not age-dependent and exhibits very low intersubject variability \cite{howarth1988does}. 
Chromatic aberration introduces a refractive difference for the eye across the wavelength range from $400$ to $700\ nm$, amounting to $2.2\ D$, which might seem considerable \cite{roorda2002humanvisual}. 
However, the impact of chromatic aberration is mitigated by the eye's wavelength-dependent sensitivity: over 70\% of luminous energy is concentrated within a defocus range of less than $0.25\ D$ on either side of focus.

Another possible explanation is related to the effect of the filtering of blue light  on achromatic and chromatic channels. For Kinney et al. the reduction of the opponent components of the two systems could result in a larger physiologic response\cite{kinney_reaction_1983}.

\section{Methods}
\subsection{Visual Acuity Measurement}
For the assessment of visual acuity, participants were instructed to read the optotypes presented on the standard ETDRS chart from a distance of 4 meters. The ETDRS logMAR chart, derived from the framework proposed by Bailey and Lovie, integrates the guidelines outlined by the U.S. National Academy of Sciences–National Research Council (NAS-NRC).

The chart is structured with five letters per row, ranging in size from $+1.0$ to $-0.30$ logMAR at the specified distance of $4$ meters. 
Participants were tasked with identifying each letter sequentially until they made an error in identifying a full row. At that point, the test was concluded, and visual acuity was computed using the methodology elucidated by Ferris et al\cite{ferris1996standardizing}

\subsection{Pelli Robson Contrast Sensitivity Chart}
The Pelli Robson chart was utilized to measure contrast sensitivity\cite{pelli1988design}. 
The subject stood at a distance of $1$ meter from the chart  where the letters subtend 2.85 degrees of visual angle\cite{njeru2021effect}. This chart displays letters arranged in two groups of three (two ``triplets'') per row, with decreasing contrast. 
Contrast remained constant within triplets, and each triplet differed from the previous one by $-0.15 \log_{10}$ Units. 
Testing concluded when the subject failed to correctly identify 2 out of the 3 letters in a triplet, and the log CS score was then recorded.

\subsection{Lighting}
\begin{figure}[htbp]
    \centering
    \includegraphics[width=\linewidth]{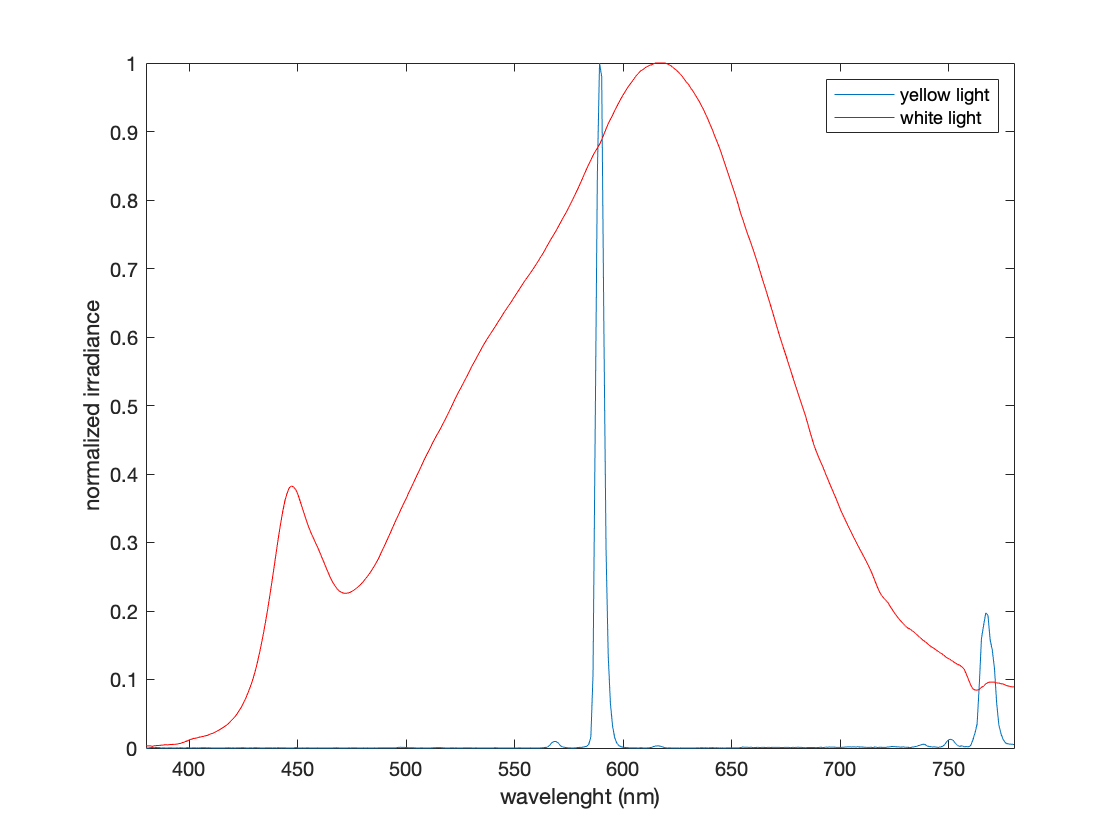}
    \caption{Spectral irradiance of the yellow lamp (blue line) and of the white lamp (red line). The maximum value of irradiance is normalized to $1$ to facilitate a comparison}
    \label{fig:spectra}
\end{figure}

The measurements for Visual Acuity (VA) and Contrast Sensitivity (CS) were conducted in two distinct locations: one situated within the exhibition space ``Olafur Eliasson: nel tuo tempo'', referred to hereafter as the ``yellow room'', and the other within the library of Palazzo Strozzi, denoted as the ``white room''.
In the yellow room the lighting was fixed, because is part of the exhibition and it was relized using low pressure sodium vapor lamps. In the white room a white LED lamp was used and the distance from the chart was set to reproduce the same luminance obtained in the yellow room. 
The arrangement of the lamp was set in order to avoid reflections or dis-homogeneity on the chart.
The luminance levels on the charts were determined using a digital light meter, registering a value of $580\ cd/m^2$. 
Notably, this measurement significantly exceeds the luminance recommendations outlined in the ETDRS protocol, which suggests a range of $80$ to $320\ cd/m^2$\cite{kaiser2009prospective}.
The difference in the lighting of the two rooms was the spectrum of the lamp: in the yellow room we had an almost monochromatic lamp (Fig.\ref{fig:spectra}).

\subsection{Participants}
\begin{figure}[htbp]
    \centering
    \includegraphics[width=\linewidth]{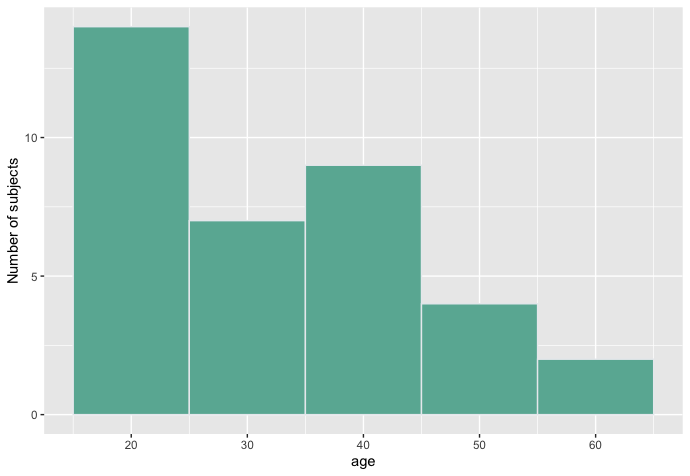}
    \caption{Histogram showing the age distribution of participants.}
    \label{fig:ageDistribution}
\end{figure}

A group of 36 participants consisted of employees of Palazzo Strozzi Fundation. All voluntereed. The age distribution is in Fig.\ref{fig:ageDistribution}.

\section{Results}
\begin{figure}[htbp]
    \centering
    \includegraphics[scale=0.55]{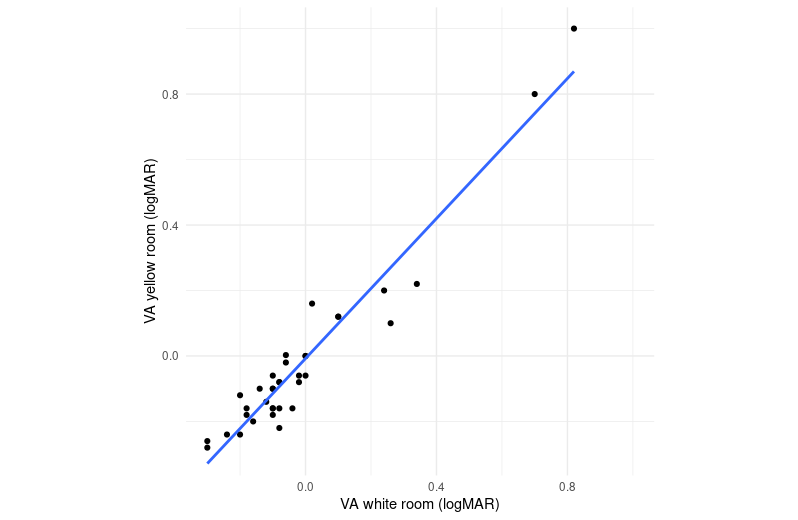}
    \caption{Scatter plot for visual acuity measurements (black dot=experimental data, line=best fit)}
    \label{fig:VAcomparison}
\end{figure}

In Figure \ref{fig:VAcomparison}, we present a scatter plot illustrating the comparison between visual acuity measurements taken in the yellow room and those in the white room. The best-fit equation, derived using the least squares method, is represented as $y = (1.07 \pm 0.05)x + (-0.008 \pm 0.01)$. The coefficient of determination $(R^2)$ for this model ,is $0.93$, indicating a robust correlation between the variables.

\begin{figure}[htbp]
    \centering
    \includegraphics[width=\linewidth]{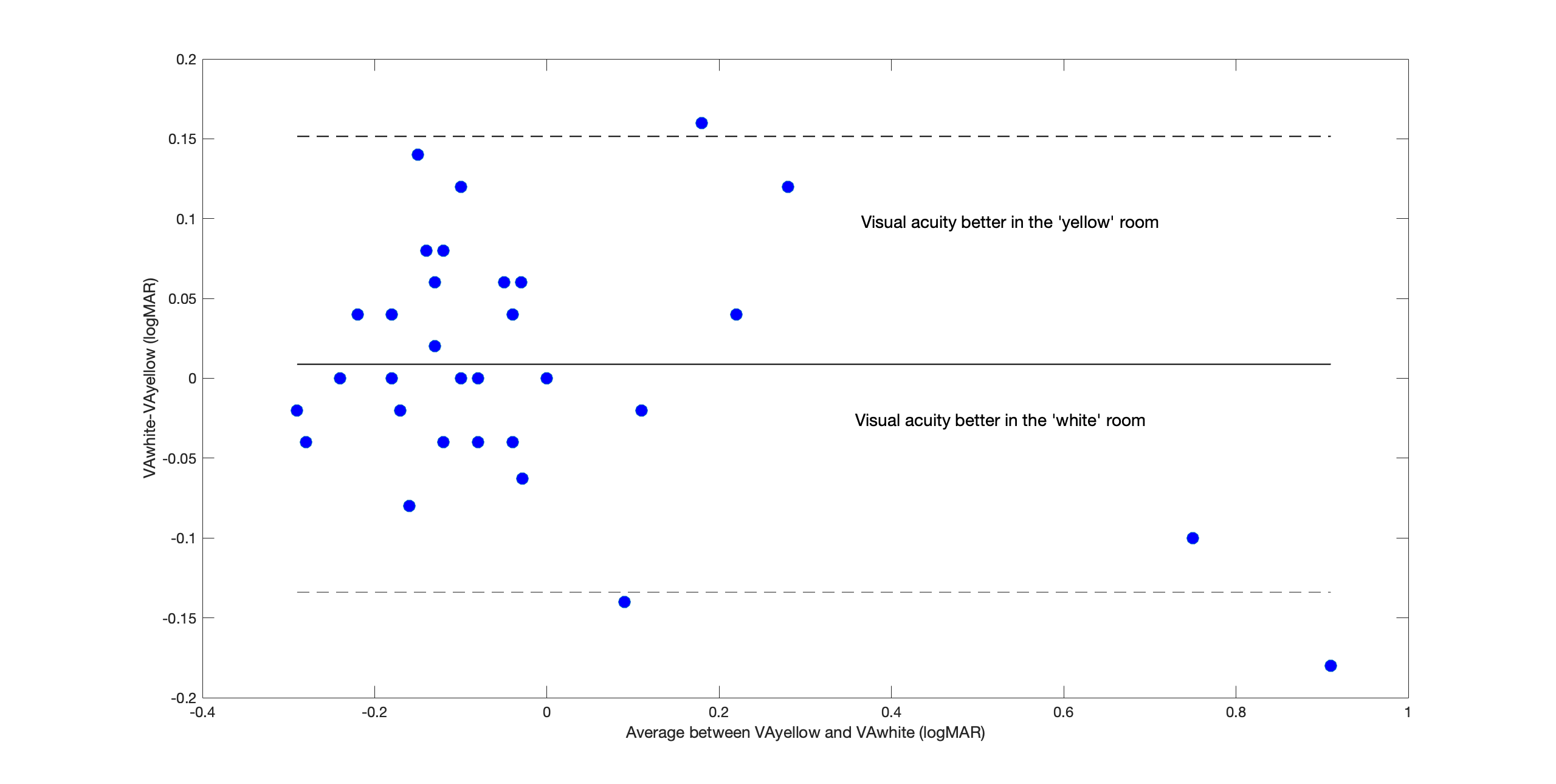}
    \caption{Bland Altman plot for visual acuity}
    \label{fig:VABlandAltman}
\end{figure}

In Figure \ref{fig:VABlandAltman}, we present the Bland-Altman plot\cite{bland1986statistical} depicting the difference between visual acuity measurements in the white room and the yellow room on the $y$-axis. 
The observed bias is minimal, with a value of $0.009$, which is considered negligible from a clinical perspective.
The $p$-value resulting from a $t$-test comparing the two measurements is notably elevated $(p=0.473)$, affirming the absence of significant differences in visual acuity between the two lighting conditions.

\begin{figure}[htbp]
    \centering
    \includegraphics[scale=0.55]{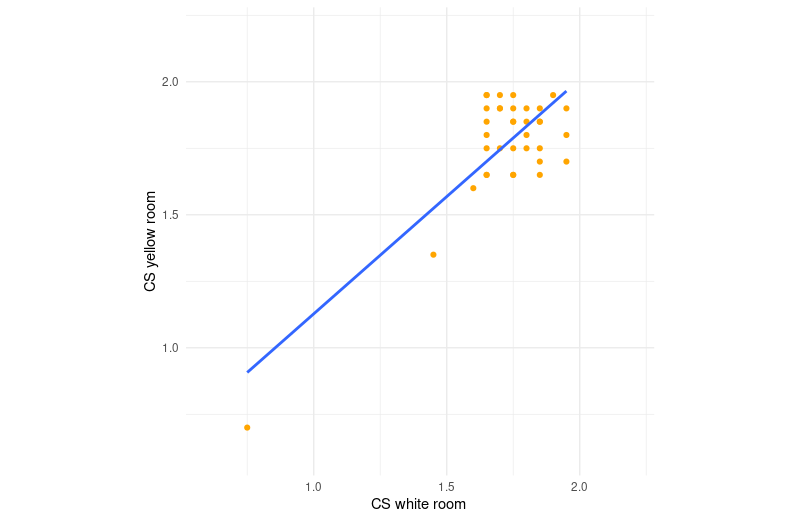}
    \caption{Scatter plot for contrast sensitivity measurements (dot=experimental data, line=best fit)}
    \label{fig:CSlinearregression}
\end{figure}

We extended our analysis to include contrast sensitivity (CS) measurements. To begin, we showcase a scatter plot (Fig.\ref{fig:CSlinearregression}) that visually captures the comparison between CS measurements obtained in both the yellow and white rooms. 
The best-fit equation, determined through the application of the least squares method, is expressed as $y = (0.88 \pm 0.12)x + (0.25 \pm 0.21)$. 
Notably, the coefficient of determination $(R^2)$ for this model is $0.62$, signifying a correlation that is comparatively weaker than that observed in VA measurements.

\begin{figure}[htbp]
    \centering
    \includegraphics[width=\linewidth]{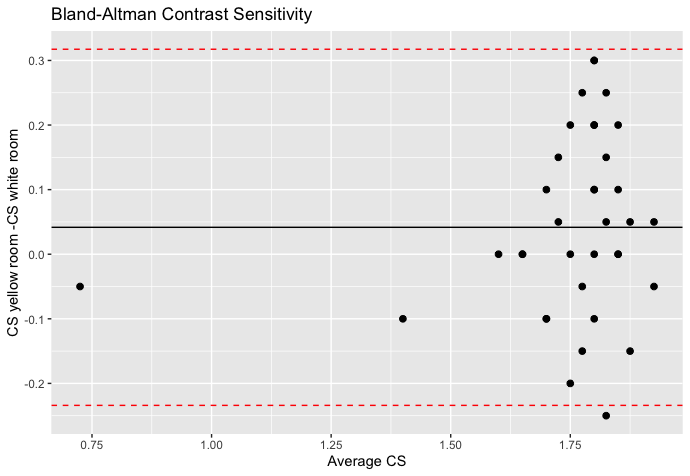}
    \caption{Bland Altman plot for contrast sensitivity}
    \label{fig:CSBlandAltman}
\end{figure}

The Bland-Altman plot is depicted in Figure \ref{fig:CSBlandAltman}. 
In this scenario, the bias, assessed as the mean of $CSyellow-CSwhite$, is $+0.04$. This indicates a superior contrast sensitivity in the yellow room, a finding that may hold clinical significance.
For contrast sensitivity (CS), the $p$-value obtained from a $t$-test for the two results is noticeably smaller compared to visual acuity (VA). 
Specifically, it is $p=0.08$, which, although not less than $0.05$, could potentially signal that a statistically significant result might be achievable with a larger and more diverse sample, encompassing variations in age and ophthalmic conditions.

\section{Conclusion}
The idea behind the study originated from the observation that many individuals visiting the exhibition claimed to ``see better'' when they were in the yellow room. 
The data presented here need to be explored further. 
While, on one hand, concerning visual acuity, we can assert that there are no differences, neither positive nor negative, when exposed to monochromatic lighting, the situation regarding contrast sensitivity is more complex. The bias obtained as the average of differences between the two situations is clinically non-negligible, and the $p$-value suggests that a more extensive study might be worthwhile. 
In addition, we used a Pelli Robson chart usage distance of 1 meter, as specified for its standard use. 
However, this leads to low spatial frequencies where the effect of monochromatic light may be less pronounced\cite{watson2013formula}. 
It might be interesting to repeat the measurements with higher spatial frequencies (e.g., 10 cycles per degree), where the difference could be more noticeable\cite{watson2013formula}.

It is evident that the study has several limitations. Firstly, it is crucial to note that the tests (both the ETDRS chart and the Pelli-Robson chart) were used in much higher illumination conditions than those for which they were validated. Additionally, the condition prevents the use of monitor-based tests (since the monitor's self-illumination would interfere with ambient lighting), which could allow for more accurate psychophysical methods.

Nevertheless, the ``laboratory'' provided by Olafur Eliasson's exhibition enabled measurements not easily achievable in a traditional laboratory.

\section{Acknowledgments}
The authors would like to express their gratitude to the Palazzo Strozzi Foundation, and in particular to Martino Margheri, Head of Educational Activities for Universities, Academies, and Special Projects, for their willingness to allow the measurements to take place within the exhibition, as well as to all the individuals who participated in the data collection. 
Additionally, they would like to thank Stefano Cavalieri for the valuable discussions.


\printbibliography
\end{document}